\documentclass[10pt,conference,pdfa]{IEEEtran}
\IEEEoverridecommandlockouts
\usepackage{graphicx} 
\usepackage{booktabs} 
\usepackage[pdfa]{hyperref}
\usepackage{multirow}
\usepackage{algorithmic}
\usepackage{graphicx}
\usepackage{textcomp}
\usepackage[table]{xcolor}
\def\BibTeX{{\rm B\kern-.05em{\sc i\kern-.025em b}\kern-.08em
    T\kern-.1667em\lower.7ex\hbox{E}\kern-.125emX}}
\usepackage[table]{xcolor} 
\usepackage{tikz} 
\usepackage{float}
\usepackage{xspace}
\usepackage{amsmath,amssymb,amsfonts} 
\usepackage[group-separator={,}]{siunitx}
\usepackage[nomessages]{fp} 
\usepackage[a-2b]{pdfx}

\title{Does the Tool Matter? Exploring Some Causes of Threats to Validity in Mining Software Repositories}
\author{
    \IEEEauthorblockN{Nicole Hoess}
    \IEEEauthorblockA{\textit{Technical University of} \\
    \textit{Applied Sciences Regensburg} \\
    Regensburg, Germany \\
    \href{mailto:nicole.hoess@othr.de}{nicole.hoess@othr.de}}
    \and
    \IEEEauthorblockN{Carlos Paradis}
    \IEEEauthorblockA{\textit{No Affiliation} \\
    Sunnyvale, CA, USA \\
    \href{mailto:cvas@acm.org}{cvas@acm.org}}
    \and
    \IEEEauthorblockN{Rick Kazman}
    \IEEEauthorblockA{\textit{University of Hawaii at Mānoa}\\
    Honolulu, HI, USA \\
    \href{mailto:kazman@hawaii.edu}{kazman@hawaii.edu}}
    \and
    \IEEEauthorblockN{Wolfgang Mauerer}
    \IEEEauthorblockA{\textit{Technical University of} \\
    \textit{Applied Sciences Regensburg} \\
    \textit{Siemens AG, Technology} \\
    Regensburg/Munich, Germany \\
    \href{mailto:wolfgang.mauerer@othr.de}{wolfgang.mauerer@othr.de}}
}

\newcommand{\eg}{\emph{e.g.}\xspace}

\newcommand{\etal}{\emph{et al.}\xspace}

\definecolor{lfd}{HTML}{E69F00}
\newcommand{\maxnum}{1.00}
\newlength{\maxlen}
\newcommand{\databar}[2][orange]{%
  \settowidth{\maxlen}{\maxnum}%
  \addtolength{\maxlen}{2.6\tabcolsep}%
  \FPeval\result{round(#2/\maxnum:4)}%
  \rlap{\color{gray!15}\hspace*{-0.1\tabcolsep}\rule[-0.15\ht\strutbox]{\maxlen}{1.2\ht\strutbox}}%
  \rlap{\color{lfd!70}\hspace*{-0.1\tabcolsep}\rule[-0.15\ht\strutbox]{\result\maxlen}{1.2\ht\strutbox}}%
  \makebox[\dimexpr\maxlen-\tabcolsep][r]{#2}%
}

\begin{document}

\maketitle

\begin{abstract}
    Software repositories are an essential source of information for software engineering research on topics such as project evolution and developer collaboration.
    Appropriate mining tools and analysis pipelines are therefore an indispensable precondition for many research activities.
    Ideally, valid results should not depend on technical details of data collection and processing. It is, however, widely acknowledged that mining pipelines are complex, with a multitude of implementation decisions made by tool authors based on their interests and assumptions. This raises the questions if (and to what extent) tools agree on their results and are interchangeable.
    In this study, we use two tools to extract and analyse ten large software projects, quantitatively and qualitatively comparing results and derived data to better understand this concern. We analyse discrepancies from a technical point of view, and adjust code and parametrisation to minimise replication differences. Our results indicate that despite similar trends, even simple metrics such as the numbers of commits and developers may differ by up to 500\%. We find that such substantial differences are often caused by minor technical details. We show how tool-level and data post-processing changes can overcome these issues, but find they may require considerable efforts.
    We summarise identified causes in our lessons learned to help researchers and practitioners avoid common pitfalls, and reflect on implementation decisions and their influence in ensuring obtained data meets explicit and implicit expectations. Our findings lead us to hypothesise that similar uncertainties exist in other analysis tools, which may limit the validity of conclusions drawn in tool-centric research.
\end{abstract}

\section{Introduction} \label{introduction}

In software projects, social entities, such as persons or groups, interact with technical artefacts, including modules, files and functions. The relationships between these two groups can be described by socio-technical networks constructed from software repositories, which capture a project's entire history of collaboration~\cite{joblin_developer_2015}. Temporal socio-technical networks can represent a project's state at any point in time. This information can help managers, new contributors and others to gain a comprehensive understanding of the project, facilitating, for example, the identification of the most suitable contact persons for addressing specific issues, such as a bug in a software module.
The analysis of collaboration networks has become a common tool in understanding evolutionary aspects of organisational principles, primarily based on open-source software (OSS) projects~\cite{herbold_systematic_2021, bock_synchronous_2021, hunsen_fulfillment_2020, joblin_evolutionary_2017, joblin_developer_2015, bock_measuring_2021, joblin_classifying_2017, bock_automatic_2023, joblin_hierarchical_2023, tamburri_evolving_2021, tamburri_exploring_2021, catolino_gender_2019, almarimi_learning_2020, de_stefano_splicing_2020, lambiase_good_2022, palomba_predicting_2021, palomba_beyond_2021, stefano_impacts_2022, eken_empirical_2021, mauerer_search_2022, picha_towards_2017, joblin_how_2022, tamburri_canary_2020}. Patterns of interest include \cite{herbold_systematic_2021} the formation of developer communities \cite{joblin_developer_2015}, the emergence of (non-)hierarchical team structures~\cite{joblin_hierarchical_2023}, and the identification of core developers~\cite{joblin_classifying_2017, bock_automatic_2023}. Socio-technical factors can also aid in predicting quality aspects, including bugs, software failures and future project success~\cite{joblin_how_2022} as well as in optimising circumstances for collaboration~\cite{bird_empirical_2011} to increase developer productivity~\cite{xuan_building_2014} and reduce development time and costs.

To facilitate these analyses, the field of mining software repositories (MSR) offers a plethora of methods and tools to extract collaboration and communication data from a variety of sources, including version control systems (VCS), mailing lists, issue trackers and chats~\cite{tymchuk_collaboration_2014}. A series of processing stages, such as matching multiple developer identities, bot detection, file and temporal filtering, may be employed to obtain a baseline data set. To construct the collaboration network, contributors are initially linked to artefacts, mailing list threads, issues or chat messages to which they contributed. A bipartite projection eliminates the source nodes from the bipartite network, instead connecting developers who contributed to them. Alternatively, a temporal projection links developers according to the temporal order of their contributions to a source. Edge weights may reflect the strength of collaboration between two developers. In co-change networks, weights can be calculated from the number of lines of code (LoC) changed in a file, function or module~\cite{joblin_how_2022, joblin_classifying_2017, joblin_developer_2015}.

Although existing tools and scripts perform similar steps of this complex pipeline, they differ in their underlying assumptions, algorithms, implementations and interpretations~\cite{paradis_building_2022}. This raises the question of whether the results 
of previous studies would have been the same if a different tool or analysis pipeline had been employed. Studies on the promises and perils of mining version control data have identified a number of factors that can potentially lead to inaccurate analysis results, but also identified and proposed mechanisms and methods to enhance consistency. To date, there are only a few studies that quantify the effects of these implementation details on analysis results and their interpretation. In particular, there is a lack of replication studies for developer networks~\cite{herbold_systematic_2021}.

Our objective is to quantify the impact and threats to validity of employing different mining tools on version control systems and utilising this baseline data in subsequent developer network analyses. To this end, we conduct a replication study to measure the similarity of the VCS data and collaboration networks mined and constructed by two independently developed tools, namely Codeface~\cite{codeface} and Kaiaulu~\cite{kaiaulu}.

Our results indicate significant uncertainties due to unknown assumptions. While overall trends largely correspond, we show that under specific circumstances, such as limitations in file filtering and identity matching, minor implementation and configuration details may have substantial impact. For instance, the percentage of developer identities jointly found by both tools may not exceed 60\% for some subject projects and even most active developers can be missed. Changing tool parametrisation and implementing post-processing could overcome parts of these issues in our case. However, not all discrepancies can be eliminated by such means, and would require major re-engineering efforts. 
Users are likely unaware of the existence and magnitude of such limitations and could derive wrong conclusions, decisions and best practices or make predictions based on inaccurate ground truth data when trusting their pipeline.
This paper deliberately focuses on the technical details responsible for this issue.

In summary, we contribute (1) quantitative insights into the impact of different mining tool implementations on the characteristics of the baseline data, (2) a qualitative analysis of the causes of these differences and (3) actionable learnings from and for the adjustment and replication of a tool to obtain outcomes as close as possible to the desired baseline data.
\section{Related Work} \label{related-work}

\subsection{Replication Studies}
The limited number of available replication studies is considered a threat to validity in empirical software engineering~\cite{basili_building_1999,robles_replicating_2010}. In general, advances in reproducibility engineering, such as dedicated platforms~\cite{ghezzi_replicating_2013, trautsch_addressing_2018} and archived, self-contained reproduction packages bundling artefacts and pipelines~\cite{mauerer_beyond_2022}, contributed significantly to the improvement of study reproducibility in the last years~\cite{gonzalez-barahona_revisiting_2023}. MSR studies yet remain challenging, as artefacts and implementations are often not published at all, only in parts, or in form of a diverse~\cite{liang_can_2024, trautsch_addressing_2018}, possibly unusable~\cite{gonzalez-barahona_revisiting_2023}, set of scripts and tools. However, replication studies offer the potential to increase confidence in findings of previous studies~\cite{shull_role_2008}, extend or complement their results~\cite{gonzalez-barahona_reproducibility_2012} and increase the impact of the field ~\cite{liang_can_2024}. For example, Dinh-Trong~\etal~\cite{dinh-trong_freebsd_2005} compare the development process of the Apache project investigated in the original work to the one of FreeBSD to enhance understanding of OSS development processes. 
Bock~\etal~\cite{bock_synchronous_2021} refine the results of a study on synchronous development, which associates developer collaboration and communication with productivity ~\cite{xuan_building_2014}, by transferring the method to a different data set and abstraction level. While replication studies are more common in MSR areas such as defect and bug detection ~\cite{mahmood_reproducibility_2018, di_penta_relationship_2020}, their availability is limited in the context of developer networks ~\cite{herbold_systematic_2021}.

\subsection{Validity Studies}

Pitfalls in MSR that can distort results have been addressed by several influential studies. Bird~\etal~\cite{bird_promises_2009} addressed possible traceability issues. Kalliamvakou~\etal~\cite{kalliamvakou_promises_2014} extend this study to include peculiar properties of GitHub. Flin~\etal~\cite{flint_pitfalls_2022} studied pitfalls when working with time-based Git data and provide guidance to overcome them. Saarimäki~\etal~\cite{saarimaki_towards_2022} investigate the perils of disregarding time dependencies in MSR studies. They propose methods from time series analysis and present preliminary results of their impact. Nia~\etal~\cite{nia_validity_2010} explore the impact of pitfalls in e-mail network construction, resulting in missing edges or edges out of temporal order. Their results show that metrics such as node centrality are stable despite the changes in topology. Meneely~\etal~\cite{meneely_socio-technical_2011} study developer perception of collaboration and team structure expressed by developer network edges and metrics. From a higher level perspective, Siegmund~\etal~\cite{siegmund_views_2015} studied experts' views on the \emph{correct} application of empirical methods in software engineering, finding that there is no consensus on the prioritisation of internal or external validity in the field.

\subsection{Method and Tool Comparisons}

In technical debt detection, Lefever~\etal compared the results found by various commercial and open-source tools, finding that they disagree even for very common, basic measures such as LoC \cite{lefever_lack_2021}. To the best of our knowledge, there are no similar tool comparisons for developer networks. However,
differentiating aspects are highlighted by comparisons of specific methods implemented as part of 
MSR pipelines. Goeminne~\etal~\cite{goeminne_comparison_2013} compare heuristics for developer identity matching, 
finding that exact substring matching performs best. Amreen~\etal~\cite{amreen_alfaa_2020} propose a novel machine learning technique for identity matching and demonstrate its superiority to state-of-the-art methods.
Bertoncello~\etal~\cite{bertoncello_pull_2020} investigate the influence of using commits or pull-requests to distinguish core and casual contributors in a software project, concluding that pull-requests are more accurate for measuring contributions. Joblin~\etal~\cite{joblin_developer_2015, joblin_classifying_2017} investigate differences in co-change collaboration network construction, community detection and core developer classification metrics, along with an evaluation of correspondence with the real perception of developers. Several authors study the agreement of developer networks~\cite{tymchuk_collaboration_2014, panichella_how_2014} and communities~\cite{bock_measuring_2021} resulting from data extracted from different communication and collaboration sources. Tymchuk~\etal~\cite{tymchuk_collaboration_2014} find that the combination of channels is essential to obtain a comprehensive view of a project. While related studies focus on specific analysis stages in isolation, we aim to explore the impact and propagation of, possibly unknown, differences in two distinct tool implementations along the entire analysis pipeline.

\section{Replication Study Design}\label{study-design}

\begin{figure}[htbp]
    \centering
    \includegraphics[width=0.3\textwidth]{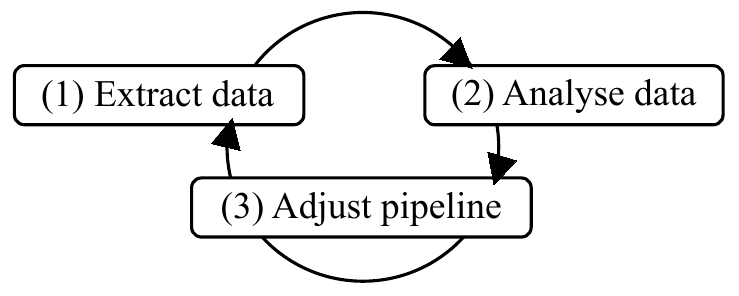}
    \caption{Overview about our approach: We mined repositories using a basis (Codeface) and replication (Kaiaulu) tool, and compared derived data. Depending on similarities and differences, we adjusted configuration parameters and post-processing methods in the replication tool. We iterated until the closest replication was achieved.}
    \label{fig:method}
\end{figure}

Our replication is based on an iterative process visualised in Fig.~\ref{fig:method}. We began with tool configurations of Codeface and Kaiaulu used in previous studies, and ran the tools on ten large software repositories to analyse the results and evaluate their agreement. Based on the identified differences and once knowing their causes, we adapted the pipeline for Kaiaulu and began the next iteration to achieve a closer replication to Codeface. For the sake of clarity, the individual iterations are not distinguished in the following sections. 

\vspace{-0.5em}
\begin{figure}[!ht]
    \centering
    \includegraphics[width=0.45\textwidth]{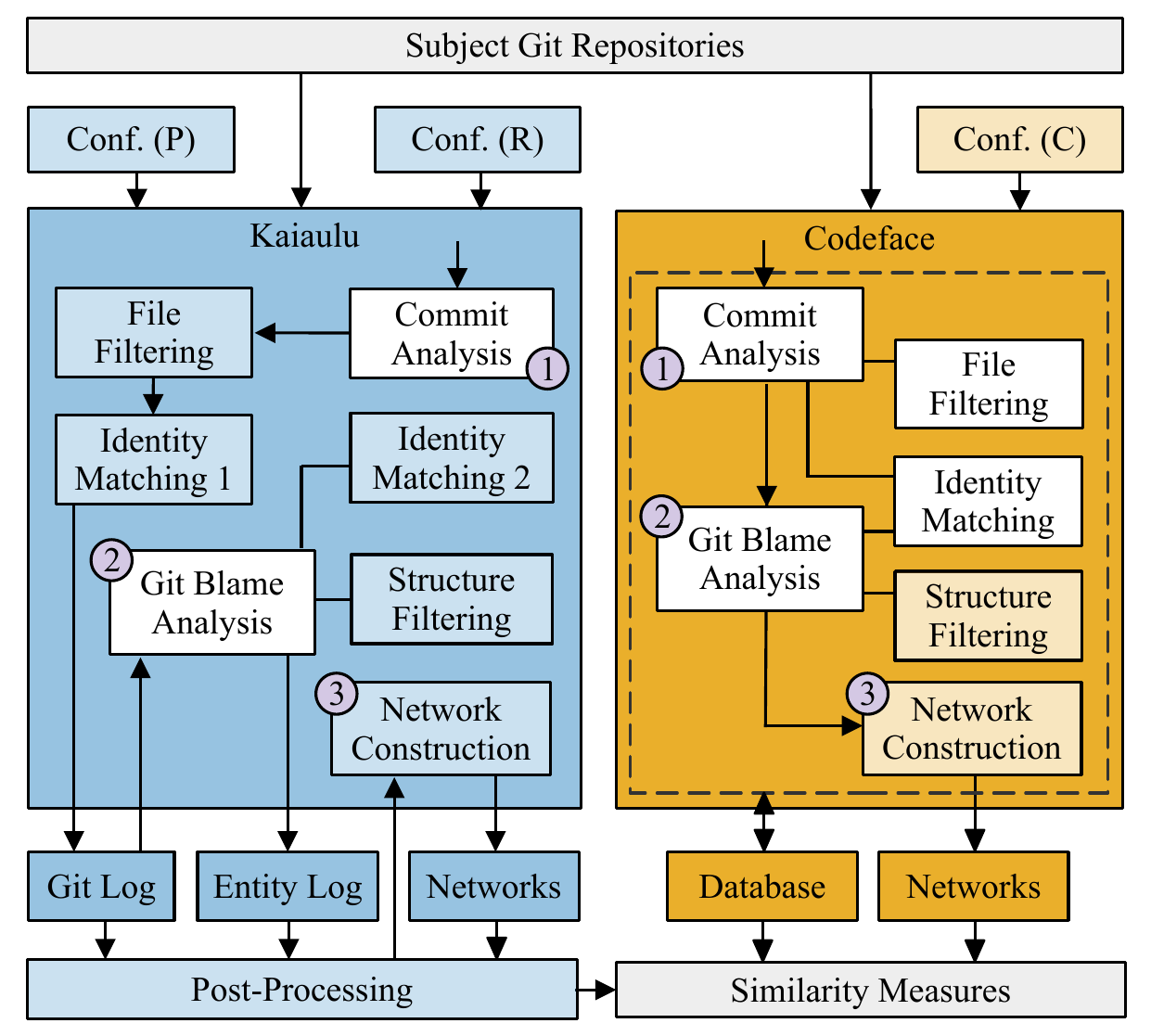}
    \caption{Informal overview of structural components of the mining tools Codeface and Kaiaulu. Although both tools perform the same analysis steps, their interaction and data structure differ.}
    \label{fig:pipeline}
\end{figure}

\subsection{Research Questions}
\textbf{RQ1:} \emph{To what extent can we observe discrepancies in the data obtained from independent MSR tools?} Each tool makes assumptions on how to extract and process data, which can be influenced by configuration parameters. Implicit assumptions and implementation decisions often remain hidden from tool users. We are interested in quantifying the impact of these uncertainties 
on the results.

\textbf{RQ2:} \emph{Which factors are responsible for the discrepancies?} Assuming we find discrepancies in RQ1, we are interested in technical causes and inherently different tool assumptions.

\textbf{RQ3:} \emph{To what extent and with what effort can a tool be adapted to yield results comparable to those produced by another tool?} Based on RQ2, we evaluate how adjusting configuration and implementation achieves the closest replication of baseline data extracted by the original tool.

Fig.~\ref{fig:pipeline} provides an overview of the analysis pipeline.

\subsection{Subject Selection}

Our data set consists of ten relevant and active OSS projects. For our selection, we take into account previous studies relying on Codeface or Kaiaulu and drawing conclusions based on these repositories~\cite{joblin_developer_2015, bock_synchronous_2021, hunsen_fulfillment_2020, joblin_evolutionary_2017, bock_measuring_2021, bock_automatic_2023,  joblin_hierarchical_2023, catolino_gender_2019, mauerer_search_2022, joblin_how_2022, paradis_analyzing_2024}. We consider different application domains, programming languages, ages of the project and project and team sizes. Tab.~\ref{tab:subjects} provides an overview of the subject characteristics. To avoid a tool bias, commits and developers are extracted directly via git, without processing such as identity matching. The primary language and LoC are determined via \textsc{cloc}~\cite{adanial_cloc}.

\begin{table*}[htb]
    \rowcolors{2}{gray!15}{white}
    \centering
    \caption{Descriptive statistics of our subject projects\label{tab:subjects}}
    \begin{tabular}{lccrrrcr}
        \toprule
         Project & Domain & Primary Language & Commits & Developers & LoC & Start & End\\ \midrule
         \textsc{Camel} & Middleware & Java & \num{68222} & \num{1352} & \num{2252114} & 2007-03-19 & 2024-01-11 \\
         \textsc{Django} & Web Framework & Python & \num{32348} & \num{2897} & \num{637711} & 2005-07-13 & 2024-01-23 \\
         \textsc{GTK} & GUI & C & \num{78899} & \num{1587} & \num{1383746} & 1997-11-24 & 2024-01-16 \\
         \textsc{Jailhouse} & Hypervisor & C & \num{3006} & \num{75} & \num{50593} & 2013-10-20 & 2023-01-10 \\
         \textsc{OpenSSL} & Secure Communication Library & C & \num{34691} & \num{1069} & \num{861736} & 1998-12-21 & 2024-01-21 \\
         \textsc{PostgreSQL} & DBMS & C & \num{57500} & \num{55} & \num{1708628} & 1996-07-09 & 2024-01-11 \\
         \textsc{QEMU} & Emulator, Virtualiser & C & \num{109894} & \num{2376} & \num{1972799} & 2003-02-18 & 2024-01-10 \\
         \textsc{RStudio} & IDE & Java, C++ & \num{41745} & \num{194} & \num{1004561} & 2010-12-07 & 2024-01-25 \\
         \textsc{Spark} & Data Analytics Engine & Scala & \num{39503} & \num{2815} & \num{1377476} & 2010-03-29 & 2024-01-11 \\
         \textsc{Wine} & Compatibility Layer & C & \num{168818} & \num{1762} & \num{5251550} & 1993-06-29 & 2024-01-10 \\
         \bottomrule
    \end{tabular}
\end{table*}

\subsection{Data Extraction}

We chose Codeface as the basis tool because it has been used in numerous studies between 2015 and 2023~\cite{joblin_developer_2015, bock_synchronous_2021, hunsen_fulfillment_2020, joblin_evolutionary_2017, bock_measuring_2021, joblin_classifying_2017, bock_automatic_2023, joblin_hierarchical_2023, tamburri_evolving_2021, tamburri_exploring_2021, catolino_gender_2019, almarimi_learning_2020, de_stefano_splicing_2020, lambiase_good_2022, palomba_predicting_2021, palomba_beyond_2021, stefano_impacts_2022, eken_empirical_2021, mauerer_search_2022, picha_towards_2017, joblin_how_2022}, \cite{tamburri_canary_2020}. We decided to use Kaiaulu as replication tool because it is relatively new with fewer studies available \cite{paradis_building_2022, paradis_analyzing_2024, paradis_socio-technical_2024, mumtaz_analyzing_2022, mumtaz_preliminary_2022}. Kaiaulu is partly inspired by Codeface \cite{paradis_building_2022} and, to the best of our knowledge, the only tool implementing \emph{all} required stages of the MSR pipeline to analyse developer collaboration in the same way as Codeface. We set up both by cloning the public, actively maintained repositories for Codeface~\cite{codeface} and Kaiaulu~\cite{kaiaulu} and following the installation instructions. 

To conduct the analyses, both tools require configuration files for the tool itself and for each project analysis run. Codeface enables its users to configure a time window size in months or specific release ranges for temporal analyses. We chose a window size of three months as proposed in Refs.~\cite{joblin_classifying_2017} and \cite{bock_measuring_2021}. In addition, Codeface offers three modes for commit analysis and network construction, which measure contributions and collaboration strengths based on files, features or code structures. For our study, we choose the structure-based \emph{proximity} mode, which is also available in a similar form with Kaiaulu's \emph{entity} mode, parsing functions and classes.

Kaiaulu's API offers a high degree of freedom for the analysis pipeline. We built on the API to define scripts for the analysis. During our study, we identified a number of decisions that influence results. Therefore, we use two distinct configurations, \emph{prior} and \emph{replication}. \emph{Prior} refers to the configuration we set up based on prior studies conducted with Kaiaulu \emph{before} the first exploration of the baseline data for RQ1. \emph{Replication} is adjusted and complemented with additional post-processing derived from RQ2 \emph{after} iteratively inspecting the mined data to most closely replicate Codeface's outputs and answer RQ3.

To ensure a fair comparison, both MSR tools use identical time windows. Codeface's time slicing starts and ends by commit timestamps instead of actual time windows, implying that a commit break of six months would not result in two empty time windows, but a larger time window ending with the commit after the break. To replicate this behaviour in Kaiaulu, we implement a new time slicing, enabling us to specify the exact window limits used by Codeface. We also add new configuration options to split time intervals by either author timestamp (prior) or committer timestamp (replication).

In Fig.~\ref{fig:pipeline}, the lighter coloured areas represent configurable parts of the tool pipelines. In Codeface, commit analysis, subsequent code structure extraction and network construction depend on the choice of analysis mode and are executed in parallel, triggered by a single interface. Depending on the configuration, the results stored in Codeface's database tables will differ. In Kaiaulu, each step during and subsequent to commit analysis is optional and uses independent configuration parameters to be specified by the user, such as file endings to filter and code structure tags to search for.
Codeface decides internally on which parser, file filters and tags to use.
In Kaiaulu's prior configuration, we filtered the same file endings and language tags as proposed in existing configurations from prior Kaiaulu studies~\cite{paradis_building_2022, paradis_analyzing_2024}, while in the replication configuration, we filtered the same file endings and language structure tags as proposed for Codeface.

Kaiaulu's identity matching is optionally performed at multiple occasions, always requiring a specification of the table and identity columns. Codeface performs identity matching as proposed by Bird~\etal~\cite{bird_mining_2006} when parsing commits across all identity columns (author and committer) and stores the unified results. We implemented similar cross-column and cross-table identity matching in Kaiaulu's replication configuration.

\subsection{Developer Network Construction}

In function-based developer networks, the collaboration strength of each pair of developers is calculated from the number of changed source lines of code \(|sloc|\) reported in the git blame data of a revision. For each commit from developer \(b\), Codeface iterates over any modified functions or code structures. For each function, it considers all commits from the collaborating developer \(a\), who previously modified this function. For each of those previous commits, Codeface adds the sum of newly contributed \(|sloc_{b,c_i}|\) from \(b\) and the remaining \(|sloc_{a,c_j}|\) of the previous commit from \(a\) to the edge weight. Kaiaulu provides several edge weight schemes for the function-based analysis mode. The scheme implemented and used for prior Kaiaulu studies also sums up \(|sloc|\), but, instead of adding the newly contributed \(|sloc_{b,c_i}|\) to each of the previous commits of \(a\), only adds them once. Therefore, we introduced a new edge weight scheme to replicate the nested sum from Codeface in Kaiaulu's replication configuration.

For two developers \(d_1\) and \(d_2\), the edge weight \(\omega_{d_2,d_1}\) is given by the sum of their contributions measured in source lines of code \(sloc\) at time of commit \(\tau_{d,t}\) as described in Eq.~(\ref{eq:codeface}) for Codeface (\(c\)) and Eq.~(\ref{eq:kaiaulu-prior}) for Kaiaulu (\(k\)) with prior configuration. \(\delta\) filters commits of \(d_1\) contributing to \(f\).
\begin{align} 
    \omega_{d_2,d_1,c} =& 
    \sum_{f \in F}\
        \Bigl\{\
            \sum_{i=0}^{n}\ 
                \Bigl[\,
                    \sum_{j=0}^{i-1}\,
                        \Bigl(|sloc_{d_2}(\tau_{d_2, i-1}, f)|\, + \label{eq:codeface}\\
                        &(|sloc_{d_1}(\tau_{d_1,j}, f)|\Bigr) 
                    \,\cdot\,\delta\bigl(|sloc_{d_1}(\tau_{d_1,j}, f)|\bigr)
                \Bigr]
        \Bigl\}\nonumber
\end{align}
\begin{align}
    \omega_{d_2,d_1,k} =& 
    \sum_{f \in F}\
        \Bigl\{\
            \sum_{i=0}^{n}\ 
                \Bigl[\,
                    |sloc_{d_2}(\tau_{d_2,i-1}, f)|
                    + \label{eq:kaiaulu-prior}\\
                    &\sum_{j=0}^{i-1}
                        |sloc_{d_1}(\tau_{d_1,j}, f)| 
                    \,\cdot\,\delta\bigl(|sloc_{d_1}(\tau_{d_1,j}, f)|\bigr)
                \Bigr]
        \Bigl\}\nonumber
\end{align}

Our comparison uses count-based metrics for number of identified commits, changed files, changed entities and developer identities over time. Time series similarity is calculated by normalised compression distance~\cite{cilibrasi_clustering_2005}, measuring the distance of the compression of two vectors individually and concatenated. Also, dynamic time warping identifies the optimal warping path aligning elements of the sequences with minimal distance~\cite{berndt_using_1994}. We include Spearman's rank correlation coefficient as intuitive measure, knowing that it can be misleading for time series due to dependent values~\cite{dean_dangers_2016}. To measure the similarity of the developer networks constructed by both tools, we calculated their density, mean non-zero edge weight and graph edit distance as a notion of required edit operations to transform one graph into the other.

\section{Results}

To present results and address research questions, 
space constraints require limiting visualisations to representative samples; full data is available in the \href{https://doi.org/10.5281/zenodo.14091455}{reproduction package}.

\subsection{Comparison of Baseline and Derived Data}

\begin{figure*}[htb]
    \includegraphics[width=\textwidth]{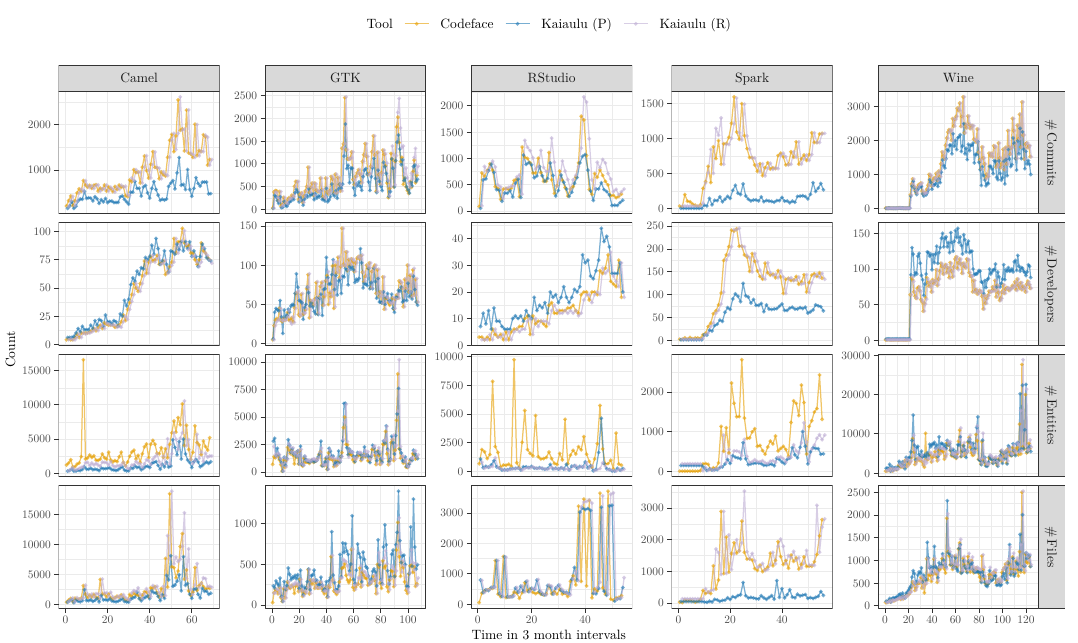}\vspace*{-1em}
    \caption{Time series of count-based metrics calculated based on the git log extracted by Codeface and Kaiaulu with prior (P) configuration and Kaiaulu with replication (R) configuration. Lines are plotted with a small horizontal offset to visualise overlapping lines. We see similar trends for both tools, but notice that the extent of discrepancies is project- and configuration-specific. The most prominent differences can be observed for project Spark.}
    \label{fig:git-ts}
\end{figure*}

\begin{figure*}[htb]
    \includegraphics[width=\textwidth]{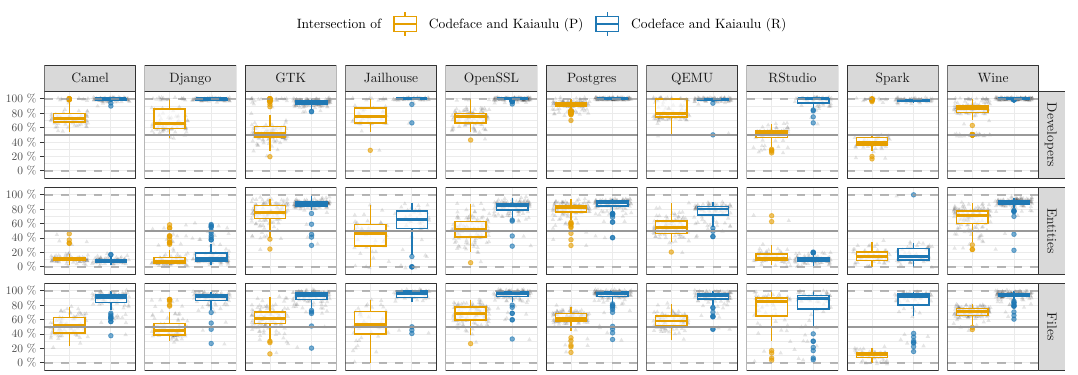}\vspace*{-1em}
    \caption{Jointly identified files, entities and developers, considering prior (P) and replication (R) configurations over the entire VCS history. Again, we observe that the magnitude of discrepancies depends on the project, the considered metric and the tool configuration.}
    \label{fig:git-overlap}
    \vspace{-0.75em}
\end{figure*}

To address RQ1, we visualise the number of commits, active developers, entities and files found by each tool configuration in the defined time intervals in Fig.~\ref{fig:git-ts}. In Tab.~\ref{tab:git-ts}, we quantitatively evaluate the visual trends, comparing the time series' normalised compression distance (NCD), dynamic time warping (DTW) distance and Spearman's rank correlation coefficient (Cor.). We show the tools' agreement on jointly identified files, entities and developer identities in Fig.~\ref{fig:git-overlap}.

\begin{table*}[htbp]
\centering
\caption{Time series similarity for Codeface and Kaiaulu with prior (P)
               and replication (R) configuration measured by Normalised
               Compression Distance (Ncd), Dynamic Time Warping (Dtw) 
               and Spearman's Rank Correlation Coefficient (Cor). For Ncd and Dtw, 
               values close to zero (gray bar) are desired;
               for correlation, values close to one (orange bar) 
               indicate higher similarity. \label{tab:git-ts}}
\centering
\begin{tabular}[t]{clrrrrrrrrrrrr}
\toprule
\multirow{17}{*}{\rotatebox[origin=c]{90}{Codeface/Kaiaulu (P)}}
& & \multicolumn{3}{c}{\textbf{Commits}} & \multicolumn{3}{c}{\textbf{Files}} & \multicolumn{3}{c}{\textbf{Entities}} & \multicolumn{3}{c}{\textbf{Developers}} \\
\cmidrule(l{3pt}r{3pt}){3-5} \cmidrule(l{3pt}r{3pt}){6-8} \cmidrule(l{3pt}r{3pt}){9-11} \cmidrule(l{3pt}r{3pt}){12-14}
\textbf{Tools} & \textbf{Project} & \textbf{NCD} & \textbf{DTW} & \textbf{Cor} & \textbf{NCD} & \textbf{DTW} & \textbf{Cor} & \textbf{NCD} & \textbf{DTW} & \textbf{Cor} & \textbf{NCD} & \textbf{DTW} & \textbf{Cor}\\
\midrule
 & Camel & \databar{0.86} & \databar{0.21} & \databar{0.92} & \databar{0.87} & \databar{0.19} & \databar{0.92} & \databar{0.89} & \databar{0.24} & \databar{0.91} & \databar{0.69} & \databar{0.08} & \databar{0.91}\\
 & Django & \databar{0.83} & \databar{0.17} & \databar{0.96} & \databar{0.86} & \databar{0.23} & \databar{0.91} & \databar{0.87} & \databar{0.24} & \databar{0.69} & \databar{0.73} & \databar{0.08} & \databar{0.69}\\
 & GTK & \databar{0.88} & \databar{0.18} & \databar{0.93} & \databar{0.87} & \databar{0.22} & \databar{0.91} & \databar{0.88} & \databar{0.18} & \databar{0.89} & \databar{0.68} & \databar{0.26} & \databar{0.89}\\
 & Jailhouse & \databar{0.67} & \databar{0.23} & \databar{0.95} & \databar{0.68} & \databar{0.27} & \databar{0.85} & \databar{0.68} & \databar{0.34} & \databar{0.88} & \databar{0.56} & \databar{0.27} & \databar{0.88}\\
 & OpenSSL & \databar{0.86} & \databar{0.12} & \databar{0.98} & \databar{0.86} & \databar{0.11} & \databar{0.96} & \databar{0.87} & \databar{0.18} & \databar{0.88} & \databar{0.78} & \databar{0.07} & \databar{0.88}\\
 & Postgres & \databar{0.89} & \databar{0.31} & \databar{0.75} & \databar{0.89} & \databar{0.22} & \databar{0.82} & \databar{0.89} & \databar{0.22} & \databar{0.91} & \databar{0.48} & \databar{0.08} & \databar{0.91}\\
 & QEMU & \databar{0.90} & \databar{0.11} & \databar{0.97} & \databar{0.87} & \databar{0.14} & \databar{0.98} & \databar{0.88} & \databar{0.22} & \databar{0.88} & \databar{0.75} & \databar{0.06} & \databar{0.88}\\
 & RStudio & \databar{0.82} & \databar{0.23} & \databar{0.94} & \databar{0.81} & \databar{0.20} & \databar{0.66} & \databar{0.80} & \databar{0.33} & \databar{0.71} & \databar{0.66} & \databar{0.17} & \databar{0.71}\\
 & Spark & \databar{0.81} & \databar{0.21} & \databar{0.88} & \databar{0.80} & \databar{0.32} & \databar{0.78} & \databar{0.80} & \databar{0.26} & \databar{0.83} & \databar{0.78} & \databar{0.11} & \databar{0.83}\\
 & Wine & \databar{0.89} & \databar{0.09} & \databar{0.99} & \databar{0.90} & \databar{0.18} & \databar{0.90} & \databar{0.92} & \databar{0.20} & \databar{0.84} & \databar{0.68} & \databar{0.07} & \databar{0.84}\\
\midrule
 & Mean \(\mu\) & \databar{0.84} & \databar{0.18} & \databar{0.93} & \databar{0.84} & \databar{0.21} & \databar{0.87} & \databar{0.85} & \databar{0.24} & \databar{0.84} & \databar{0.68} & \databar{0.12} & \databar{0.84}\\
 & Std. Dev. \(\sigma\) & \databar{0.06} & \databar{0.06} & \databar{0.07} & \databar{0.06} & \databar{0.06} & \databar{0.09} & \databar{0.07} & \databar{0.05} & \databar{0.07} & \databar{0.09} & \databar{0.08} & \databar{0.07}\\
\midrule
\multirow{10}{*}{\rotatebox[origin=c]{90}{Codeface/Kaiaulu (R)}}
 & Camel & \databar{0.82} & \databar{0.02} & \databar{1.00} & \databar{0.85} & \databar{0.12} & \databar{0.98} & \databar{0.86} & \databar{0.20} & \databar{0.92} & \databar{0.49} & \databar{0.02} & \databar{0.92}\\
 & Django & \databar{0.53} & \databar{0.05} & \databar{1.00} & \databar{0.83} & \databar{0.16} & \databar{0.98} & \databar{0.86} & \databar{0.14} & \databar{0.91} & \databar{0.43} & \databar{0.01} & \databar{0.91}\\
 & GTK & \databar{0.63} & \databar{0.10} & \databar{0.98} & \databar{0.86} & \databar{0.12} & \databar{0.98} & \databar{0.89} & \databar{0.06} & \databar{0.94} & \databar{0.58} & \databar{0.03} & \databar{0.94}\\
 & Jailhouse & \databar{0.11} & \databar{0.00} & \databar{1.00} & \databar{0.61} & \databar{0.14} & \databar{0.98} & \databar{0.68} & \databar{0.21} & \databar{0.92} & \databar{0.18} & \databar{0.03} & \databar{0.92}\\
 & OpenSSL & \databar{0.12} & \databar{0.00} & \databar{1.00} & \databar{0.85} & \databar{0.07} & \databar{0.98} & \databar{0.87} & \databar{0.04} & \databar{0.98} & \databar{0.29} & \databar{0.01} & \databar{0.98}\\
 & Postgres & \databar{0.07} & \databar{0.00} & \databar{1.00} & \databar{0.89} & \databar{0.12} & \databar{0.93} & \databar{0.88} & \databar{0.17} & \databar{0.94} & \databar{0.10} & \databar{0.00} & \databar{0.94}\\
 & QEMU & \databar{0.75} & \databar{0.03} & \databar{1.00} & \databar{0.89} & \databar{0.10} & \databar{0.98} & \databar{0.87} & \databar{0.13} & \databar{0.97} & \databar{0.25} & \databar{0.00} & \databar{0.97}\\
 & RStudio & \databar{0.78} & \databar{0.12} & \databar{0.97} & \databar{0.77} & \databar{0.21} & \databar{0.67} & \databar{0.80} & \databar{0.34} & \databar{0.64} & \databar{0.58} & \databar{0.08} & \databar{0.64}\\
 & Spark & \databar{0.56} & \databar{0.10} & \databar{0.96} & \databar{0.80} & \databar{0.16} & \databar{0.93} & \databar{0.76} & \databar{0.25} & \databar{0.88} & \databar{0.65} & \databar{0.01} & \databar{0.88}\\
 & Wine & \databar{0.07} & \databar{0.00} & \databar{1.00} & \databar{0.90} & \databar{0.07} & \databar{0.98} & \databar{0.92} & \databar{0.06} & \databar{0.97} & \databar{0.39} & \databar{0.01} & \databar{0.97}\\
\midrule
 & Mean \(\mu\) & \databar{0.44} & \databar{0.04} & \databar{0.99} & \databar{0.83} & \databar{0.13} & \databar{0.94} & \databar{0.84} & \databar{0.16} & \databar{0.91} & \databar{0.39} & \databar{0.02} & \databar{0.91}\\
 & Std. Dev. \(\sigma\) & \databar{0.30} & \databar{0.05} & \databar{0.01} & \databar{0.08} & \databar{0.04} & \databar{0.09} & \databar{0.07} & \databar{0.09} & \databar{0.09} & \databar{0.17} & \databar{0.02} & \databar{0.09}\\
\bottomrule
\end{tabular}
\end{table*}

\textbf{\textit{Commits:}} Fig.~\ref{fig:git-ts} shows that the results from both tools describe similar trends in the time series of commits. The DTW distance and Spearman's rank correlation coefficient presented in Tab.~\ref{tab:git-ts} confirm this quantitatively, although NCD appears more sensitive. Codeface usually reports significantly higher numbers of commits than Kaiaulu with prior configuration. The extent of discrepancies depends on the project and is particularly evident for Camel and Spark. In one of Spark's analysis time intervals, Codeface extracted 1593 commits, while Kaiaulu only identified 331. The difference in commits is crucial, because each subsequent step in the analysis pipelines, shown in Fig.~\ref{fig:pipeline}, relies on them. 

\textbf{\textit{Files:}} Aligned with the commit time series, the file time series in Fig.~\ref{fig:git-ts} indicates similar trends for Codeface and Kaiaulu. However, despite Codeface usually identifying more commits, Kaiaulu appears to find more files in certain project contexts such as GTK and Wine. The percentage of jointly identified files compared by their names is shown in Fig.~\ref{fig:git-overlap} over all time intervals. Again, we observe a strong project-dependent intersection, with values below \num{20}\% for Spark and up to \num{90}\% for RStudio when considering the prior configuration.

\textbf{\textit{Entities:}} The time series of changed entities shown in Fig.~\ref{fig:git-ts} do not allow any clear conclusions to be drawn. While we see very similar curves for the GTK and Wine projects, the similarity is not as pronounced for Spark and almost non-existent for RStudio. We can also recognise this trend from the metrics presented in Tab.~\ref{tab:git-ts}. Although correlation is high, the DTW distances of \num{0.34} for Jailhouse and \num{0.33} for RStudio are the worst in the table. The jointly identified entity names show a similar picture as the time series: in Fig.~\ref{fig:git-overlap}, we can see a high intersection of \num{60}-\num{80}\% in the GTK and Wine projects, while the intersection in Camel, RStudio and Spark is low.

\textbf{\textit{Developers:}} Developer time series in Fig.~\ref{fig:git-ts} differ to a similar extent as the commit time series. Surprisingly, the direction of discrepancies is reverse for some projects (\eg, Camel, RStudio and Wine). This indicates that Kaiaulu with prior configuration detects more developers than Codeface, although it identifies less commits. The time series similarity measures presented in Tab.~\ref{tab:git-ts} do not capture this phenomenon.
The agreement on developer identities shown in Fig.~\ref{fig:git-overlap} gives a better intersection than for files and entities. However, agreement on developer identities can again drop below \num{60}\%. 

Discrepancies in the four dimensions are not always related, but may have different causes. For instance, for Camel, we observe a high intersection of \num{60}-\num{80}\% in developers, while the agreement on identified entities is below \num{20}\%.

\begin{figure}[!ht]
    \vspace*{-0.4em}\includegraphics[width=0.49\textwidth]{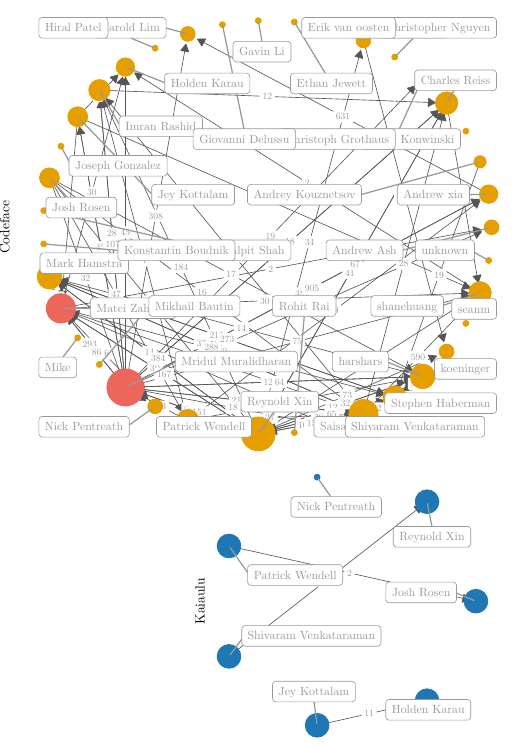}\vspace*{-2.1em}
    \caption{Developer networks constructed by Codeface (orange) and Kaiaulu with prior configuration (blue) for Spark from March--June '13. Two very active developers missed by Kaiaulu are marked red. The discrepancies in networks are due to Kaiaulu not supporting the Scala language.}
    \label{fig:network-comparison}
    \vspace{-1em}
\end{figure}

\textbf{\textit{Developer Networks:}} To determine similarities between constructed networks, we compute graph edit distance, network density and mean (non-zero) edge weight for each time interval (see Tab.~\ref{tab:network-similarity}).
Intuitively, graph edit distance is large for projects with several thousand developers (\eg, QEMU and Django), and small for projects with several dozen to hundred developers (\eg, Jailhouse, Postgres and RStudio). As for time series, Spark represents an outlier with a graph edit distance twice as high as the second-largest observation. This indicates that discrepancies from the baseline data do propagate to derived data. Tab.~\ref{tab:network-similarity} also shows that networks derived by Kaiaulu are in general denser than by Codeface, while edge weights from the latter considerably exceed the former in its prior configuration. For QEMU, the maximum edge weight found by Kaiaulu is \num{11960}, while Codeface reports a value of \num{261362}---almost \num{22} times larger. 
Manual inspection reveals that several developers were missed by Kaiaulu; especially two of the most active developers according to Codeface in one time interval of Spark (see Fig.~\ref{fig:network-comparison}). Anomalies such as duplicate developer identities persist throughout
the networks.

\begin{table*}[htbp]
\centering
\caption{Graph similarities determined by Codeface (C) and 
               Kaiaulu (K) with prior (P) and replication (R) configuration.
               \label{tab:network-similarity}}
\centering
\rowcolors{2}{gray!15}{white}
\begin{tabular}[t]{lrrrrrrrrrrrrr}
\toprule
\multicolumn{1}{c}{\textbf{}} & \multicolumn{4}{c}{\textbf{Graph Edit Distance}} & \multicolumn{3}{c}{\textbf{Density}} & \multicolumn{6}{c}{\textbf{Edge Weight}} \\
\cmidrule(l{3pt}r{3pt}){2-5} \cmidrule(l{3pt}r{3pt}){6-8} \cmidrule(l{3pt}r{3pt}){9-14}
\rowcolor{white}\multicolumn{1}{c}{\textbf{}} & \multicolumn{2}{c}{\textbf{C/K(P)}} & \multicolumn{2}{c}{\textbf{C/K(R)}} & \multicolumn{1}{c}{\textbf{C}} & \multicolumn{1}{c}{\textbf{K(P)}} & \multicolumn{1}{c}{\textbf{K(R)}} & \multicolumn{2}{c}{\textbf{C}} & \multicolumn{2}{c}{\textbf{K(P)}} & \multicolumn{2}{c}{\textbf{K(R)}} \\
\cmidrule(l{3pt}r{3pt}){2-3} \cmidrule(l{3pt}r{3pt}){4-5} \cmidrule(l{3pt}r{3pt}){6-6} \cmidrule(l{3pt}r{3pt}){7-7} \cmidrule(l{3pt}r{3pt}){8-8} \cmidrule(l{3pt}r{3pt}){9-10} \cmidrule(l{3pt}r{3pt}){11-12} \cmidrule(l{3pt}r{3pt}){13-14}
\textbf{Project} & \textbf{Mean} & \textbf{Max} & \textbf{Mean} & \textbf{Max} & \textbf{Mean} & \textbf{Mean} & \textbf{Mean} & \textbf{Mean} & \textbf{Max} & \textbf{Mean} & \textbf{Max} & \textbf{Mean} & \textbf{Max}\\
\midrule
Camel & \num{119} & \num{399} & \num{140} & \num{474} & \num{0.17} & \num{0.17} & \num{0.21} & \num{1009} & \num{474484} & \num{298} & \num{167093} & \num{1633} & \num{3036609}\\
Django & \num{162} & \num{533} & \num{236} & \num{744} & \num{0.13} & \num{0.20} & \num{0.24} & \num{510} & \num{188873} & \num{47} & \num{9431} & \num{230} & \num{356944}\\
GTK & \num{141} & \num{534} & \num{164} & \num{682} & \num{0.02} & \num{0.07} & \num{0.11} & \num{642} & \num{185640} & \num{120} & \num{31874} & \num{706} & \num{339516}\\
Jailhouse & \num{5} & \num{15} & \num{4} & \num{11} & \num{0.25} & \num{0.34} & \num{0.30} & \num{539} & \num{9608} & \num{26} & \num{664} & \num{124} & \num{4943}\\
OpenSSL & \num{63} & \num{351} & \num{86} & \num{442} & \num{0.29} & \num{0.33} & \num{0.44} & \num{759} & \num{170390} & \num{130} & \num{38087} & \num{1471} & \num{1192273}\\
Postgres & \num{47} & \num{231} & \num{48} & \num{229} & \num{0.37} & \num{0.36} & \num{0.43} & \num{960} & \num{63799} & \num{247} & \num{19578} & \num{1677} & \num{1316173}\\
QEMU & \num{374} & \num{849} & \num{544} & \num{1436} & \num{0.20} & \num{0.18} & \num{0.20} & \num{4469} & \num{261362} & \num{73} & \num{11960} & \num{951} & \num{10429560}\\
RStudio & \num{33} & \num{175} & \num{46} & \num{246} & \num{0.39} & \num{0.59} & \num{0.97} & \num{7019} & \num{157248} & \num{969} & \num{31959} & \num{92754} & \num{4765887}\\
Spark & \num{848} & \num{1951} & \num{873} & \num{2068} & \num{0.13} & \num{0.07} & \num{0.10} & \num{336} & \num{63907} & \num{92} & \num{27929} & \num{347} & \num{159043}\\
Wine & \num{251} & \num{877} & \num{647} & \num{2083} & \num{0.03} & \num{0.02} & \num{0.12} & \num{4200} & \num{512257} & \num{61} & \num{108895} & \num{158} & \num{352091}\\
\bottomrule
\end{tabular}
\end{table*}

\textbf{\textit{Answering RQ1:}} We observe similar 
evolutionary trends for count statistics of commits, developers, 
modified files and entities. The extent of (dis-)agreement
however depends on measure, project context and tool configuration. In case of the Spark project, for instance, we observe an absolute percentage difference of up to 500\% in the number of commits. Substantial discrepancies in nodes and edge weights of derived developer networks confirm that baseline data differences propagate to high-level
analyses, and merit close scrutiny.

\subsection{Investigation of Discrepancies}

To address RQ2, we analyse differences between baseline and derived data, and summarise findings. First, we filter instances detected by one but missed by the other tool. We then manually review a random subset of instances using VCS history, Codeface's database, Kaiaulu's git and entity log tables, and adjacency matrices of the developer networks.

\textbf{\textit{Commits:}} The majority of commits missed by Kaiaulu 
concerned files that use programming languages not specified in the 
prior configuration. Unlike Codeface, commit extraction in Kaiaulu is 
not generic, but requires to explicitly specify file suffixes to be 
analysed. In prior studies, Kaiaulu's configurations considered 
suffixes .c, .cc, .ccp, .java, .js, .py, and .r. Commits to language 
files such as .scala, primary language in Spark, and commits to 
non-code files such as .html, .md and .xml were missed by Kaiaulu, 
but captured by Codeface. Other commits missed by Kaiaulu indicate in their commit message that they were 
cherry-picked, refactorings or tests. This discrepancy can be 
partially attributed to
the exclusion of directories that supposedly contain tests and 
code examples in the prior configuration. Another reason could be that 
Codeface's parsing of the git blame output from step (2) in Fig.~\ref{fig:pipeline}
iteratively updates commits 
in its database if a line has been changed by a commit not yet 
captured.
Such updates of the commit table are not performed by Kaiaulu. After parsing the git log with Perceval in step (1), Fig.~\ref{fig:pipeline}, captured commits may only be reduced, but no more data is added to the table.

The difference data set also includes commits missed by Codeface and 
captured by Kaiaulu. These commits were usually merge commits of pull requests or feature branches. However, not all merge commits were missed 
by Codeface. We also note that, occasionally, commits appear in 
different time intervals in Codeface and Kaiaulu. Kaiaulu's prior 
configuration excludes the end time stamp from a time interval, while 
Codeface includes it. Further discrepancies arise as Codeface splits 
time intervals based on the committer timestamps, while Kaiaulu uses 
author timestamps, meaning that time series peaks could be shifted at the transition between two intervals.

\textbf{\textit{Files:}} Inspecting file difference data reveals discrepancies
caused by schema differences. Kaiaulu stores commits multiple times in
a table, once for each modified file. 
This gives a direct relation between commit and file. Codeface, however, stores each 
commit only once in its commit database table. In proximity analysis mode, file names are
recorded in the additional commit dependency table filled \emph{after} file and code structure filtering 
during git blame analysis. Consequently, a file is only recorded 
if a code 
structure (\eg, function) in the file was edited. 
Thus, Codeface 
occasionally misses files found by Kaiaulu.
All files 
could be determined by a second Codeface analysis using \emph{file} mode.
However, this workaround would lead to issues such as multiple internal identifiers
for identical objects, requiring a substantial merging effort. Alternatively, we could adjust 
Codeface's database schema to store files and code structures separately. This  would 
require restructuring the entire analysis process depicted in Fig.~\ref{fig:pipeline}.

\textbf{\textit{Entities:}} Discrepancies in identified entities can result from parsing.
Codeface applies, based on file extensions, Doxygen, C-Tags or an internal SQL parser
to extract structural elements. Kaiaulu uses C-Tags uniformly.
However, similar to file filtering, Kaiaulu requires to configure tags.
In prior studies, Kaiaulu only parsed function and class tags for C, C++, Java, Python and R. 
Codeface, when opting for the C-Tags parser internally, parses a default set of tags to detect structures such as functions, enumerations, namespaces, typedefs, macros, and more for \emph{any} language. More tags for special entities (\eg, constructors) are defined for individual languages such as Go.

\textbf{\textit{Developers:}} Developers missed by Kaiaulu 
emphasise the relevance of file filtering, because many of these developers contributed to files with neglected extensions in the prior configuration. Comparing developer names shows that higher developer counts by Kaiaulu are caused by different means of identity matching. In Codeface, developers using multiple identities are merged \textit{across} columns and tables considering both author and committer columns. In Kaiaulu's prior studies, identities were matched only within the author column. This results in duplicate identities consisting of different sets of names and e-mail addresses of the same person.

\textbf{\textit{Developer Networks:}} Identifying causes of developer network discrepancies is complex.
For instance, filtering few selected programming languages in Kaiaulu's configuration risks that developers focusing on other languages may be under-represented.
Reconstructing the edge weight formulas \ref{eq:codeface} and \ref{eq:kaiaulu-prior} defined (implicitly) by the tool authors was especially time-consuming. As shown in Fig.~\ref{fig:pipeline}, an analysis run in Codeface passes all steps from commit analysis to network construction and large projects require
run-times of multiple days. Using Kaiaulu's mock data generator 
(650 LoC), which creates small test repositories, and running Codeface on these while adding targeted debugging statements to the code, revealed that Kaiaulu's prior edge weight scheme did not account for repeated past contributions, but only considered them once.

\textbf{\textit{Answering RQ2:}} Discrepancies arise due to interaction and aggregation of several factors: File filtering and design decisions such as the order of data extraction steps and data schema influence identified commits and, consequently, files, entities and developers. The choice (and configuration) of code structure parsers may further limit 
obtained entities. Limitations in identity matching cause major discrepancies in identified developers and derived collaboration networks. Different definitions of collaboration strength affect edge weights in the developer network.

\subsection{Adjustments for Improving Similarity}
Based on RQ2 findings, we summarise adjustments in parametrisation, code and post-processing extensions, and give effort estimates to assess
practical feasibility of the measures.

\textbf{\textit{Commits:}} To overcome the substantial difference in extracted commits, we define identical file filters (as proposed by Codeface) in Kaiaulu's replication configuration, and deactivate file path filters not implemented in Codeface. While this does not require tool changes, we face a trade-off between detected commits, files and developers in Kaiaulu: Either deactivate file filtering and keep all commits with all developers, but include many non-code files such as documentation in the file count. Alternatively, activate file filtering and miss commits to these files, including developers. In both scenarios, discrepancies 
remain. 
Addressing this issue requires either changing Codeface's data schema or Kaiaulu's, requiring major changes. A workaround is a second Kaiaulu run \emph{without} file filtering, and use this second table to calculate commit and developer metrics, but rely on the tables obtained \emph{with} file filtering for file and entity metrics. Despite the use in this study, possible inconsistencies make this method unattractive.

To address commit shifts caused by time interval splitting, we introduce new configuration options to Kaiaulu that (a) allow us to include or exclude boundaries in time windows, and (b) switch between author and committer timestamps. This leads to closely matching commit time series, as seen in Fig.~\ref{fig:git-ts}, particularly for Camel, Spark and Wine, and confirmed quantitatively by the decrease in NCD and DTW distance and the increase in correlation in Tab.~\ref{tab:git-ts}.

\textbf{\textit{Files:}} Having two versions of Kaiaulu's git log table, as previously motivated, provides closely matching file counts (see Fig.~\ref{fig:git-ts}). Time series similarity increases particularly for Spark. Tab.~\ref{tab:git-ts} confirms that especially for DTW distance. 
With the adjustments, the percentage of jointly identified files in Fig.~\ref{fig:git-overlap} increased to a median of over 90\% in all projects.

\textbf{\textit{Entities:}}
We adapted C-Tags code structure tags in Kaiaulu's replication configuration. However, using identical language-specific and generic
C-Tags as Codeface
could not increase similarity, also for previously neglected languages such as Scala, despite
slight improvements seen in Figures~\ref{fig:git-ts}, \ref{fig:git-overlap} and Tab.~\ref{tab:git-ts}.
Adding targeted debugging statements to the code reveals that the C-Tags output provides unused information (\eg, Scala functions) when using the replication configuration.\footnote{Codeface uses a library parser, Kaiaulu relies
on internal string matching. Since the line format reported by C-Tags for languages such as Scala differs from the format in the prior configuration, Kaiaulu cannot process them.}
Overcoming this limitation would require to adjust the C-Tags parsing in Kaiaulu.
Still, discrepancies could remain, given
that Codeface incorporates a Doxygen and custom SQL parser for code structure identification, and both would need to be implemented in Kaiaulu for an exact replication. As we estimate an implementation time of weeks to months, we consider such adjustments impracticable.

\textbf{\textit{Developers:}}
We introduced configuration options to include several columns in Kaiaulu's identity matching. As identity matching in Kaiaulu is still performed \emph{within} individual columns and tables, we implemented post-processing (400 LoC) to unify identities \emph{across} columns and tables.\footnote{Identity matching inserts identities from Kaiaulu's git log table and all git entity tables in the respective temporal order into a new single source of truth table. If a first match by name and e-mail address fails, we try to match based on each of their e-mail addresses. If this fails, developers are matched based on full name. Only if neither an e-mail address nor their full name matches, a new identity is created. Then, all previous identities in the git and entity tables are replaced by the respective unique identities from the identity table.} As shown in Figs.~\ref{fig:git-ts},~\ref{fig:git-overlap} and Tab.~\ref{tab:git-ts} for Kaiaulu's replication configuration, this results in a significantly improved similarity, with a median percentage of jointly identified developers of over 90\% for all projects.

\textbf{\textit{Developer Networks:}} To address different interpretations of developer collaboration strength, we implemented the edge-weight scheme identified in RQ2 
in Kaiaulu (100 LoC). Tab.~\ref{tab:network-similarity} shows higher mean edge weights compared to the prior configuration for all subject projects. In most cases, replication weights are similar to Codeface. However, Kaiaulu can lead to outliers in edge weights, such as \num{10429560} for QEMU, compared to \num{261362} with Codeface. Manually inspecting the developer networks and entities found by Kaiaulu's replication configuration identifies a high self-collaboration on test files as cause. For the same reasons, Kaiaulu 
finds a mean edge weight of \num{92754} compared to \num{7019} by Codeface for
RStudio. Thus, the new weight scheme introduced in Kaiaulu cannot 
universally increase similarity to Codeface, as likewise observed for 
graph edit distance.

Discrepancies also remain as not all language C-Tags are supported by Kaiaulu. Even though Fig.~\ref{fig:git-ts} shows that the number of developers matches closely for the replication configuration, a developer \emph{must} appear in the entity tables to be considered in network construction. Entity tables, however, only capture developers who contributed to entities with supported C-Tags; developers who only contributed in unsupported languages thus can still be missing. This also applies to the Scala developers in Spark (Fig.~\ref{fig:network-comparison}).
After the adjustments, duplicate identities were merged and the extended structure tags resulted in more edges, leading to smaller networks with higher mean density (Tab.~\ref{tab:network-similarity}). Yet, we could not reach the desired similarity to Codeface's developer networks.

Adjustments were made iteratively, as visualised in Fig.~\ref{fig:method}, as small details such as obtaining edge weight formulae required several days of effort. The required analysis re-runs took about a week. More than ten such runs were needed to achieve the presented replication configuration.

\textbf{\textit{Answering RQ3:}} For commit, developer and file metrics, we achieve very close results by adjusting tool parametrisation, implementing tool extensions and post-processing, and applying workarounds. The identified entities as basis for network construction could not be aligned more closely; this would require a major rework of one of the tools. 
\section{Lessons Learned} \label{lessons}

\textbf{\textit{Commit Analysis:}} 
Even in extracting commits from VCS histories (step (1) in Fig.~\ref{fig:pipeline}), tools may differ in terms
of filtering, aggregation and information storage. Evolutionary analyses divide the history into time intervals by multiple approaches:
Kaiaulu relies on author timestamps, while Codeface uses committer timestamps. Also time interval limits can be handled differently.
More severely, different commits and files may be captured: While Codeface stores \emph{all} commits and later filters file endings implicitly, Kaiaulu defines file filters in the configuration file in advance. Differing file path filters (\eg, for test files) may lead to more discrepancies. This is
caused since MSR tools are developed with initial intentions:
Codeface's commit analysis is motivated by constructing developer networks. When code structures such as functions are used as source, higher-level entities such as files are not relevant, and only file dependencies in which a code structure of interest has been found and edited are stored. However, an external user could assume that all files can be found in the corresponding database table, which can lead to false conclusions. 

To avoid such misinterpretation, Kaiaulu encourages data verification prior to each analysis step. 
This is essential to find anomalies such as partially matched developer identities: Kaiaulu failed to merge authors and committers correctly due to different e-mail addresses and names in both columns. Since Kaiaulu used author timestamps in all analysis steps, merging identities across columns was not required, and duplicate or inconsistent identities in both columns were not an issue. However, when expecting the behaviour of Codeface, which merges identities across tables and columns, discrepancies that strongly influence count-based developer metrics arose. Creating a unified identity table as a single source of truth for all person-related information is advisable.

Data updates may also be handled differently by the tools. While Kaiaulu parses commits using Perceval and fixes its output as a baseline for commits, Codeface performs updates of its commit table in the further course of the pipeline. For instance, during the git blame analysis, it may add previously overlooked commits from the git blame output for analysis.

\textbf{\textit{Git Blame Analysis:}} 
Code structures such as functions in step (2) in Fig.~\ref{fig:pipeline} are
determined by third-party tools (C-Tags, Doxygen, \dots) from git blame data. Our results show that choice of parser and detail configuration can strongly influence statistics. For instance, Kaiaulu relies on C-Tags, whereas Codeface uses it when Doxygen and the built-in SQL parser are not applicable. 
C-Tags configurations differ depending on programming languages and structure tags. Even though both tools have the same amount of tags available, they don't handle them the same way. First, Codeface applies a default set of tags to \emph{any} language file not filtered out in step (1) in Fig.~\ref{fig:pipeline}, while
Kaiaulu requires an explicit specification of tags. Secondly, third-party tool outputs are parsed differently.

If not explicitly communicated,
such details can easily be lost: Developers contributing code in a particular language were overlooked by Kaiaulu, which could
only be inferred from a qualitative comparison of the baseline data. This is problematic for large-scale mining without qualitative inspection. Therefore, users and tool developers should check which project properties and third-party tool settings are actually supported, and verify their data qualitatively to reduce uncertainty.

\textbf{\textit{Network Construction:}} 
Herbold~\etal~\cite{herbold_systematic_2021} report a lack of guidelines for developer social network research.
For instance, a question of interpretation arises from edge weighting. Original and replication tool consider collaboration strength as a measure of contributed LoC to an entity, but in the absence of a precise definition, the influence of past contributions remains unclear, and edge weights can vary substantially. Thus, subsequent analyses such as core developer and community detection for team structuring could lead to different results.
\section{Threats to Validity} \label{threats}

\textbf{\textit{Construct Validity:}} The comparisons in our study can be based on a wide range of data sources and metrics. Therefore, there is a risk that we draw conclusions based on the wrong choices. We reduced this risk by using established parameter choices from previous studies and by evaluating multiple metrics simultaneously. However, due to the large number of customisable parameters, we did not test all of them in detail. For example, we fixed analysis time intervals to non-overlapping three-month windows. A larger window size could lead to other results, although studies reported a minor effect \cite{mauerer_search_2022}. Also, we did not include communication data from e-mails and issue trackers, as this
exceeds the scope of the study.

We did not evaluate if our results correctly reflect reality or whether both tools produce wrong results. Earlier studies using Codeface addressed this issue through interviews with a large group of OSS developers, finding that network construction at function level \cite{joblin_developer_2015} is perceived as an accurate representation of collaboration. Thus, we consider this threat as minor.

\textbf{\textit{Internal Validity:}} We identified discrepancies between tools in an automated way, but additional manual investigations were needed to explore the root causes. Since divergences were high in absolute numbers, we could not look at each of them individually. Instead, we selected a random subset for inspection. Although we are confident that major causes of differences are identified, as including more instances did not reveal new causes, it is possible that minor edge cases remain.

\textbf{\textit{External Validity:}} Our study is based on only two MSR tools. Both were developed by different research groups, but the groups worked together at numerous prior studies and thus are assumed to have a common understanding of assumptions. While this facilitates a fair comparison, it also introduces a bias. Tools from other research groups could make fundamentally different assumptions and lead to even more divergent results. Given the limited scope of the study, the inclusion of further tools would have not allowed us to address the diverse set of MSR pipeline stages, which we prioritised to evaluate the impact and interplay of discrepancies.

Another threat is that we only analysed a set of ten subject projects. To limit the impact on generality, we considered a wide range of application areas, programming languages, project and team sizes. Further projects written in other programming languages could have led to much more varying results. For example, for project WordPress mainly written in PHP, Kaiaulu did not find any entities as PHP is not supported. We will look at more diverse repositories in future studies. 
\section{Discussion and Conclusion}

Our study shows that uncertainties in MSR pipelines 
can lead to substantial discrepancies for the extracted baseline data. Minor implementation details can have considerable consequences: Two tools sharing similar goals, methods and processing steps can output substantially different results. 

Some of the discrepancies are caused by technical limitations, such as missing language support. Other differences arise from subjective choices, for instance how strongly past collaborations should influence 
edge weights in developer networks. Many of these factors cannot be easily adjusted by changing tool parametrisation or implementing minor extensions and post-processing, but require a substantial rework of tool capabilities to
ascertain result agreement. Therefore, differences remain in our replication. It is
likely that discrepancies would be even more pronounced for other 
combinations of MSR tools not developed by collaborating groups.

Our work underscores the need for MSR replication studies 
to promote a common understanding and higher standardisation of mining processes.
To revisit the methodology, a closer investigation of the implications of identified discrepancies on the generality of conclusions drawn in empirical software engineering is needed. 
Therefore, we plan a follow-up study with a systematic literature review following the guidelines proposed by Kitchenham~\etal~\cite{kitchenham_procedures_2004, kitchenham_systematic_2013} to replicate the most influential findings in recent decades with diverse tools.

Finally, we argue that assumptions made by each tool or study should be described as fully as possible and supplemented by code. Considering these technical details in replication studies and in the design and evaluation of novel tools and pipelines can improve fairness and reliability.
We hope that the process and findings from this replication will be helpful as a reference for these efforts in the future.
\section{Reproduction Package}
For understandability and to facilitate future comparisons, we provide containerised study tools and scripts 
in our \href{https://github.com/lfd/saner2025}{GitHub repository}. A self-contained reproduction package and Docker image including OS, libraries, input and results data \cite{mauerer_beyond_2022} is available at \href{https://doi.org/10.5281/zenodo.14091455}{Zenodo}.
\vspace{0.75em}

{\parindent0pt \textbf{Acknowledgements:} WM and NH acknowledge support by the High-Tech Agenda of the Free State of Bavaria.}

\bibliography{references}
\end{document}